\documentstyle[preprint,eqsecnum,prd,aps]{revtex}
\begin{document}
\draft
%%%%%End of Preamble
%%%%Start of Text%%%%%%%%%%%%%%%%%%%%%%%%%%%%%%%%%%%%%%%%%%%%%%%%%%%%%%%
%\tightenlines
\preprint{
\vbox{\halign{&##\hfil\cr
& ANL-HEP-PR-01-22  \cr
& September, 2001    \cr
&\vspace{0.6truein}   \cr
}}}

\title{Resummation of QCD Corrections to the\\ 
$\eta_c$ Decay Rate}

\author{Geoffrey T.  Bodwin}
\address{High Energy Physics Division, Argonne National Laboratory,
Argonne, IL 60439}
\author{Yu-Qi Chen}
\address{Institute of Theoretical Physics, Academia Sinica, P.O.Box
2735, Beijing 100080 }

\maketitle

\begin{abstract}

We examine the ratio of the decay rate of the $\eta_c$ into light
hadrons to the decay rate into photons and find that most of the large
next-to-leading-order (NLO) correction is associated with running of the
strong coupling $\alpha_s$. We resum such contributions by analyzing
final-state chains of vacuum-polarization bubbles. We show that the
nonperturbative parts of the bubble chains can be absorbed into a
color-octet matrix element, once one has used contour deformations of
the phase-space integrals to cancel certain contributions. We argue that
these contributions are incompatible with the uncertainty principle.  We
also argue that perturbation theory is reliable only if one carries out
the phase-space integrations before the perturbation summation. Our
results are in good agreement with experiment and differ considerably
from those that one obtains by applying the scale-setting method of
Brodsky, Lepage, and Mackenzie to the NLO result.

\end{abstract}


\vfill \eject

\narrowtext
\section{Introduction}

In this paper we consider the decay of the heavy quarkonium $\eta_c$ into
light hadrons. According to the analysis of Ref.~\cite{bbl}, the decay
rate is given by the Nonrelativistic Quantum Chromodynamics (NRQCD)
factorization formula
\begin{equation}
\Gamma(\eta_c\rightarrow {\rm LH})=\sum_n{2\, {\rm Im}\, f_n\over 
m_c^{d_n-4}}\langle \eta_c|{\cal O}_n|\eta_c\rangle.
\label{fact-form}
\end{equation}
Here, $m_c$ is the charm-quark mass, the ${\cal O}_n$ are operators in
NRQCD with mass dimension $d_n$, and $2\, {\rm Im}\, f_n/m_c^{d_n-4}$ is
a short-distance coefficient. A similar formula applies for the
electromagnetic decay $\eta_c\rightarrow\gamma\gamma$.

In general, there are difficulties in obtaining accurate results from
the factorization formula (\ref{fact-form}). The sum is an expansion in
powers of the heavy-quark--antiquark relative velocity $v$. Only the
first few operator matrix elements in that sum are known at all, and
they are not known very accurately. The short-distance coefficients may,
in principle, be computed in perturbation theory. However, the
perturbation series is only asymptotic. In particular, it exhibits a
factorial growth that is associated with the presence of renormalon
singularities in its Borel transform \cite{beneke}. Such singularities
give rise to ambiguities in the evaluation of the perturbation series.

One can avoid some of these difficulties by considering the ratio of the
rate for $\eta_c$ into light hadrons to the rate for $\eta_c$ into two
photons. We denote this ratio by $R$. In $R$, the dependences of the
decay rates on the operator matrix elements cancel in leading order in
$v$, and the corrections of relative order $v^2$ also cancel \cite{bbl}.
These cancellations occur because, aside from an overall color factor,
the tree-level Feynman diagrams for $Q\bar Q\rightarrow gg$ and $Q\bar
Q\rightarrow\gamma\gamma$ are the same. One consequence of these
cancellations is that renormalon ambiguities associated with
initial-state virtual-gluon corrections also cancel
\cite{braaten-chen,bodwin-chen} at the level of accuracy of relative
order $v^2$. However, there is still the potential for renormalon
ambiguities associated with final-state corrections to appear.

Through next-to-leading order in $v^2$, the ratio $R$ is given by the
ratio of the short-distance coefficients of the leading-order operators
\cite{bbl}. Through next-to-leading order (NLO) in $\alpha_s$, one
obtains \cite{Barbieri:1979be,Hagiwara:1981nv}
\begin{eqnarray}
R^{\rm NLO}(\mu)=R_0(\mu)
\left[1+\left({199\over 6 }-{13\pi^2 \over 8}
-{8\over 9}n_f\right){\alpha_s(\mu) \over \pi }
+ 2\alpha_s(\mu)\beta_0 \ln {\mu^2\over 4 m_c^2}\right],
\label{r:next-leading}
\end{eqnarray}
where 
\begin{equation}
R_0(\mu)={81 C_F \alpha_s^2(\mu)\over 32 N_c\alpha^2},
\label{r0}
\end{equation}
$m_c$ is the pole mass of the charm quark, $n_f$ is the number of light
quarks ($n_f=3$ for charmonium), $\beta_0$ is the leading coefficient of
the beta function, with  $\beta_0\,=\, (33-2n_f)/(12\pi)$, and, in QCD,
$N_c=3$, and $C_F=4/3$.

Setting the factorization scale $\mu$ equal to $2m_c$ and taking
$\alpha_s(2m_c)=0.247$, one finds that
\begin{equation}
R^{\rm NLO}(2m_c)=2.1\times 10^3.
\end{equation}
This is to be compared with the world-average experimental value
\cite{PDG}
\begin{equation}
R^{\rm Exp}=3.3\pm 1.3 \times 10^{3}\,.
\end{equation}
However, the next-to-leading-order QCD correction is quite large. For
$\alpha_s(2m_c)=0.247$, it is about 1.1 times the leading-order term.
Hence, there is some question as to the reliability of the perturbation
expansion.

Furthermore, the dependence on the factorization scale is large. For
example, taking $\mu=m_c$, with $\alpha_s(m_c)=0.350$, we obtain
\begin{equation}
R^{\rm NLO}(m_c)=5.0\times 10^3,
\end{equation}
which is factor of $2.4$ larger than $R^{\rm NLO}(2m_c)$.

One popular choice of scale is that given by the
Brodsky-Lepage-Mackenzie (BLM) method \cite{BLM}. In this approach, all
terms proportional to $n_f$ are absorbed into the running coupling
$\alpha_s(\mu)$. For $R$, one obtains $\mu_{\rm BLM}\approx 0.52 m_c$
\cite{BLM}. This yields a next-to-leading-order coefficient that is only
about $0.5$ times the leading-order coefficient. However,
\begin{equation}
R^{\rm NLO}(\mu_{\rm BLM})= 9.9\times
10^3,
\label{R-BLM}
\end{equation}
which is much larger than either the theoretical result at $\mu=2m_c$
or the experimental value.

We can gain some insight into the origins of the 
large next-to-leading-order correction and the strong scale dependence 
by re-writing Eq.~(\ref{r:next-leading}) as follows:
\begin{equation}
R^{\rm NLO}(\mu)=R_0(\mu)
\left\{1+\left[\left({37\over 2 }-{13\pi^2 \over 8}\right)
+\pi\beta_0\left({16\over 3}+2\ln {\mu^2\over 4 m_c^2}\right)
\right]{\alpha_s(\mu) \over \pi}\right\}.
\label{r:next-leading-2}
\end{equation}
In the next-to-leading-order correction, the first term in square
brackets has a value of about 2.46, while the term $16\pi\beta_0/3$ has
a value of 12. Hence, we see that most of the large next-to-leading-order
correction is proportional to $\beta_0$. That is, it arises from the
one-loop corrections associated with the running of $\alpha_s$.

This suggests that we control the large next-to-leading-order correction
by resumming, to all orders in perturbation theory, the contributions
associated with the one-loop running of $\alpha_s$. Such a resummation
would have the added benefit of reducing the scale sensitivity, since
the resummed contribution would be invariant under a change of scale,
aside from two-loop corrections to the running of $\alpha_s$.
Unfortunately, there is some arbitrariness in the resummation procedure.
The contributions to the running of $\alpha_s$ that arise from quark
loops are contained entirely in chains consisting of quark
vacuum-polarization bubbles connected by gluon propagators. However, the
contributions that arise from gluon loops may, depending on the gauge,
be associated with diagrams other the gluon vacuum polarization.

One unambiguous procedure for carrying out such a resummation is the
method of ``naive non-Abelianization'' (NNA) \cite{naive-non}. In this
method, one resums the chains of quark-loop vacuum-polarization bubbles
and then accounts for the gluonic contributions by promoting the
quark-loop contribution to $\beta_0$ to the full QCD $\beta_0$. This
method is a generalization of the BLM scale-setting method. However, as
we shall see, it leads to a very different numerical result. An
alternative method for carrying out the resummation is to resum both the
quark-loop and gluon-loop vacuum-polarization bubble chains in the
background-field gauge \cite{background-field}. In this gauge, all
of the corrections associated with the running of $\alpha_s$ are
contained in the quark and gluon vacuum-polarization diagrams. It turns
out that the background-field-gauge method gives numerical results that
are close to those of the NNA method.

Vacuum-polarization insertions first appear, in relative order
$\alpha_s$, in the final-state gluon legs. Therefore, in order to resum
the leading bubble-chain corrections, we need consider only
vacuum-polarization bubble chains in the two final-state gluon legs. 
By ``leading bubble-chain corrections,'' we mean those corrections that 
contain one power of the large coefficient in the second term in square 
brackets in Eq.~(\ref{r:next-leading-2}) for each power of $\alpha_s$. 

The integrals over the final-state phase space include regions in which
the constituents of the vacuum-polarization diagrams (quarks or gluons)
have a small invariant mass. Hence, the calculation is, in principle,
sensitive to nonperturbative contributions. In the bubble-chain series,
this sensitivity may manifest itself as a factorial growth that is
characteristic of the presence of renormalon singularities in the Borel
transform of the series. 

In the case of a single chain of vacuum-polarization bubbles, we show,
by making use of a contour-deformation argument in the complex
invariant-mass plane, that the low-virtuality contributions, and, hence,
the factorial growth, cancel when one carries out the phase-space
integration before carrying out the perturbation summation. Somewhat
surprisingly, even when factorial growth in the series cancels, there is
an ambiguity that arises in the bubble-chain series: one obtains
different results, depending on whether one carries out the perturbation
summation before the phase-space integration, or {\it vice versa}. We
argue that the perturbation series is reliable only when one carries out
the phase-space integration first.

For the more general case of two bubble chains, there is a true
sensitivity to the low-virtuality region, even after contour-deformation
arguments have been applied. Indeed, the perturbation series exhibits a
factorial growth that is characteristic of the presence of renormalon
singularities in its Borel transform. We show that the low-virtuality
contributions can be identified with contributions to the decay rate
that arise from the matrix element of a color-octet NRQCD operator.
Hence, by taking into account the color-octet contribution, one can
systematically remove the low-virtuality region from the perturbative
calculation. In this analysis, contour deformation arguments are crucial
to demonstrate the cancellation of certain low-virtuality contributions
that do not correspond to matrix elements of NRQCD operators. We argue
that such contributions are inconsistent with the uncertainty principle.

The color-octet matrix element is inherently nonperturbative in nature.
We do not calculate its contribution, which is of relative order $v^3$.
Instead, we estimate the size of the contribution and treat it as an
uncertainty in the calculation.

The remainder of this paper is organized as follows. In
Sec.~\ref{sec:bubble-chain-conts}, we compute the contributions of the
vacuum-polarization bubble chains to all orders in $\alpha_s$. We
re-arrange the real and virtual contributions to obtain manifestly
infrared finite expressions. In Sec.~\ref{sec:nonpert}, we discuss the
contributions from the nonperturbative regions of small invariant mass,
first for a single bubble chain and then for two bubble chains. We
identify the nonperturbative contributions that remain after the
contour-deformation cancellation with contributions to the matrix
elements of a color-octet operator. Sec.~\ref{sec:efficient} contains a
re-arrangement of the bubble-chain contributions that is efficient for
numerical evaluation. In Sec.~\ref{sec:results}, we present our
numerical results, and in Sec.~\ref{sec:discussion} we give a summary
and a discussion of our findings.

\section{The Bubble-Chain Contributions to $R$}
\label{sec:bubble-chain-conts}

In this section, we compute the contributions to $R$ from the insertion
of any number of vacuum-polarization bubbles into the two final-state
gluons in the lowest-order Feynman diagrams for $Q\bar Q\rightarrow gg$.

These contributions are obtained by cutting the diagrams for the
heavy-quark--antiquark forward scattering amplitude through light-quark
and gluon final states in all possible ways. A cut can pass through two
vacuum-polarization bubbles, a final-state gluon and a
vacuum-polarization bubble, or two final-state gluons. We call such
contributions real-real, virtual-real, and virtual-virtual,
respectively. These contributions are separately infrared divergent, but
the Kinoshita-Lee-Nauenberg (KLN) theorem \cite{kln} ensures that the
infrared divergences cancel in the complete expression for the decay
width.

\subsection{The Effects of Bubble Chains in the Hadronic Cross Section}

In NRQCD, at leading order in $v$, the decay of a $Q\bar Q$ ${}^1S_0$
state proceeds through the color-singlet operator
\begin{equation}
{\cal O}_1({}^1S_0)=\psi^\dagger \chi \chi^\dagger\psi,
\end{equation}
where $\psi$ is the two-component Pauli spinor that annihilates the
charm quark and $\chi^\dagger$ is the two-component Pauli spinor that
annihilates the charm antiquark. According to the NRQCD factorization
formula \cite{bbl}, the contribution of this operator to the decay rate
in NRQCD is given by
\begin{equation}
\Gamma(H)
=2\,{\rm Im}\,\left[{f_1({}^1S_0)\over m_c^2}\right]
\langle H|{\cal O}_1({}^1S_0)|H\rangle,
\label{nrqcd-rate}
\end{equation}
where $H$ is a hadronic state.

We can compute the short-distance coefficient $2\,{\rm
Im}\,[{f_1({}^1S_0)/m_c^2}]$ in perturbation theory by taking the state
$H$ to be a free $Q\bar Q$ state. Then, Eq.~(\ref{nrqcd-rate}) becomes
$\Gamma(Q\bar Q)$, the rate for the annihilation of a free
heavy-quark--antiquark pair into gluons through the operator ${\cal
O}_1({}^1S_0)$.

At leading order in the heavy-quark--antiquark velocity $v$ in the
quark-antiquark center-of-mass frame, the amplitude for the decay of a
heavy quark and antiquark in a color-singlet ${}^1S_0$ state with
momenta $p_1$ and $p_2$, respectively, into two massive gluons with
momenta $k$ and $l$, polarizations $\mu$ and $\nu$, and color indices
$a$ and $b$, respectively, is given by
\begin{equation}
A_{ab}^{\mu\nu}(Q\bar Q)={1\over \sqrt{N_c}}{\rm Tr}(T_aT_b)\,2\sqrt{2}g^2\,
\epsilon^{\nu\rho\mu 0}k_\rho\,{1\over 2m_ck_0+k^2}\, ,
\label{a-gg}
\end{equation}
where $g$ is the strong coupling constant, $T_a$ is an $SU(3)$ color
matrix in the fundamental representation, and the trace is over the
indices of the $SU(3)$ matrices. In deriving Eq.~(\ref{a-gg}), we have
made use of the spin-0 projector, accurate to leading order in $v$:
\begin{equation}
\sum_{s_{1z},s_{2z}}u(p_1,s_1)\overline v(-p_2,s_2)\,
\langle s=0|s_{1}s_{2}\rangle =
{1\over 4\sqrt 2\, m_c^2}(\not\! p_1+m_c){1+\gamma_0\over 2}
\gamma_5(\not\! p_2-m_c)\,.
\end{equation}

Taking the absolute square of the quantity in Eq.~(\ref{a-gg}), inserting
vacuum-polarization bubbles in the gluon propagators, dividing by two
for Bose statistics, and integrating over the phase space, we obtain the
bubble-chain contribution to the $Q\bar Q$ decay width:
\begin{eqnarray}
\Gamma^{\rm Bub}(Q\bar Q)=&&2g^4C_F \sum_{m,n=0}^\infty
\int {d^4k\over (2\pi)^4}\theta(k_0)\int {d^4l\over (2\pi)^4}\theta(l_0)
\epsilon^{\nu\rho\mu 0}k_\rho \epsilon_{\nu\sigma\mu 0}k^\sigma
\biggl({1\over 2mk_0+k^2}\biggr)^2\nonumber\\
&&\times 2\,{\rm Im}[K^{(m)}(x)]\,
2\,{\rm Im}[K^{(n)}(y)]
(2\pi)^4\delta^4(p_1+p_2-k-l),
\label{bub-rate}
\end{eqnarray}
where
\begin{mathletters}
\begin{eqnarray}
x&&=k^2/(4m_c^2),\\
y&&=l^2/(4m_c^2).
\end{eqnarray}
\label{x-y}%
\end{mathletters}%
Aside from a polarization-tensor factor,
\begin{mathletters}%
\begin{equation}
iK^{(m)}(x)=[-i\Pi(x)]^m{i\over 4m_c^2(x+i\epsilon)}
\end{equation}
is the amplitude for the $m$th-order contribution to the bubble-chain
gluon propagator. We also define the complete propagator
\begin{equation}
K(x)=\sum_{m=0}^\infty K^{(m)}(x).
\end{equation}
\label{bub-prop}%
\end{mathletters}%
$\Pi(x)$ is defined by
\begin{equation}
\Pi_{\mu\nu}(k)=(k^2g_{\mu\nu}-k_\mu k_\nu)\Pi(x),
\label{pi-mu-nu}
\end{equation}
where $\Pi_{\mu\nu}(k)$ is the amplitude for a vacuum-polarization
bubble with external momentum $k$ and polarization indices $\mu$ and
$\nu$. We note the useful kinematic relations
\begin{mathletters}
\begin{eqnarray}
&k_0=m_c(1+x-y),&\\
&l_0=m_c(1-x+y),&\\
&{\mathbf k}^2={\mathbf l}^2=m_c^2(1+x^2+y^2-2x-2y-2xy).&
\end{eqnarray}
\label{kinematics}%
\end{mathletters}
\vfill\eject

Using Eq.~(\ref{bub-rate}) and 
\begin{equation}
\langle Q\bar Q|{\cal O}_1({}^1S_0)|Q\bar Q\rangle=2N_c,
\end{equation}
we compare the left and right sides of Eq.~(\ref{nrqcd-rate}) to
obtain the short-distance coefficient
\begin{equation}
2\,{\rm Im}\,\left[{f_1({}^1S_0)\over m_c^2}\right]={1\over 2N_c}
\Gamma^{\rm Bub}(Q\bar Q).
\end{equation}
Then, taking the hadronic state $H$ in Eq.~(\ref{nrqcd-rate}) to be the
$\eta_c$, we obtain the decay width of the $\eta_c$ into two gluons in
the bubble-chain approximation:
\begin{eqnarray}
\Gamma^{\rm Bub}(\eta_c)=&&
{g^4C_F\over N_c}\sum_{m,n=0}^\infty
\int {d^4k\over (2\pi)^4}\theta(k_0)\int {d^4l\over (2\pi)^4}\theta(l_0)
\epsilon^{\nu\rho\mu 0}k_\rho \epsilon_{\nu\sigma\mu 0}k^\sigma
\biggl({1\over 2mk_0+k^2}\biggr)^2\nonumber\\
&&\times 2\,{\rm Im}[K^{(m)}(x)]\,
2\,{\rm Im}[K^{(n)}(y)]
(2\pi)^4\delta^4(p_1+p_2-k-l) \langle\eta_c|{\cal
O}_1(^1S_0)|\eta_c\rangle.
\label{eta-width}
\end{eqnarray}

In the $\eta_c$ width into two gluons (\ref{eta-width}), we change
variables by making use of Eq.~(\ref{x-y}), which implies that
$2k_0dk_0\,\theta(k_0)=4m_c^2 dx\,\theta(x)$ and
$2l_0dl_0\,\theta(l_0)=4m_c^2 dy\,\theta(y)$, and
Eq.~(\ref{kinematics}). We use the energy-momentum-conserving
$\delta$-functions to carry out the integrations over the 3-momenta.
Then, dividing by the width into two photons,
\begin{equation}
\Gamma_{\rm EM}(\eta_c)={32\pi\alpha^2\over 81m_c^2}
\langle\eta_c|{\cal O}_1(^1S_0)|\eta_c\rangle,
\label{2-photons}
\end{equation}
we obtain the contribution to $R$:
\begin{equation}
R^{\rm Bub}=R_0\sum_{m,n=0}^\infty
\int_0^1{dx\over 2\pi}\int_0^1{dy\over 2\pi}
\,2\,{\rm Im}[4m_c^2K^{(m)}(x)]
\,2\,{\rm Im}[4m_c^2K^{(n)}(y)]\,f(x,y)\theta(1-\sqrt{x}-\sqrt{y}),
\label{R-bub}
\end{equation}
where
\begin{equation}
f(x,y)={[1-2(x+y)+(x-y)^2]^{3/2}\over (1-x-y)^2}.
\label{f-xy}
\end{equation}

In order to compute $R^{\rm Bub}$ from Eq.~(\ref{R-bub}), we need
the value of a single vacuum-polarization bubble. A standard calculation
of the light-quark bubble in dimensional regularization yields
\begin{equation}
\Pi(x) = {i \over \epsilon}
\alpha_s\beta_0
 \left(b x^{-\epsilon}-a \right) \,,
\label{vac-pol}
\end{equation}
where
\begin{mathletters}
\begin{equation}
b =\Biggl[1+\epsilon\left(
       -\gamma + \ln 4 \pi + {5\over 3}
      + \ln {\mu^2 \over 4m_c^2}\right) +O(\epsilon^2)\Biggr]
e^{i\pi\epsilon}\,,
\label{vac-pol-NNA-b}
\end{equation}
\begin{equation}
a =1+\epsilon\left(-\gamma + \ln 4 \pi + {5\over 3} + C
\right)+O(\epsilon^2)\,.
\label{vac-pol-NNA-a}
\end{equation}
\label{vac-pol-NNA}%
\end{mathletters}
In Eq.~(\ref{vac-pol}), we have renormalized the ultraviolet divergence
by subtracting the ultraviolet pole in $\epsilon$ and some associated
constants. $\epsilon$ now plays the role of an infrared regulator
($\epsilon<0$). $C$ is a subtraction-scheme-dependent constant; in
$\overline{MS}$, $C=-5/3$. We note that the phase of $b$ in
Eq.~(\ref{vac-pol-NNA-b}) and the reality of $a$ are exact to all orders
in $\epsilon$. For the light-quark bubble, $\beta_0\,=\, -2n_f/(12\pi)$.
However, in employing NNA, we promote $\beta_0$ to the QCD value
$\beta_0\,=\, (33-2n_f)/(12\pi)$.

An alternative to the NNA procedure is to compute the vacuum
polarization, including both the quark and gluon contributions, in the
background-field gauge. The motivation for this approach is that, in the
background-field gauge, the logarithmic dependence on the
renormalization scale $\mu$ is contained entirely in the
vacuum-polarization diagrams \cite{abbott}. In the background-field
gauge, the vacuum-polarization contribution still takes the form
(\ref{vac-pol}). The coefficients $a$ and $b$ are computed in
Appendix~\ref{app:background} and are given in
Eq.~(\ref{a-b-background}).

\subsection{Decomposition into Real and Virtual Contributions}

It is convenient to separate $R^{\rm Bub}$ into real-real, real-virtual,
and virtual-virtual contributions. First, we make use of the identity
\begin{equation}
{\rm Im}(AB)=({\rm Im}A)B^*+A({\rm Im}B)
\end{equation}
to identify the real and virtual parts of a single propagator:
\begin{equation}
{\rm Im}[4m_c^2K^{(m)}(x)]={1\over x-i\epsilon}
{\rm Im}\,[-i\Pi(x)]^m
-\pi\delta(x)[-i\Pi(0)]^m.
\end{equation}
Then, we substitute this result into Eq.~(\ref{R-bub}) to obtain 
\begin{equation}
R^{\rm Bub}=R_0(G_{2R}+2G_{V}G_{1R}+G_{V}^2),
\label{total-bub}
\end{equation}
where the three terms in parentheses correspond, respectively, to the
real-real, real-virtual, and virtual-virtual contributions.
\begin{equation}
G_{2R}=\sum_{m,n=1}^\infty\int_0^1 {dx\over 2\pi x}
\int_0^1 {dy\over 2\pi y} f(x,y)I_R^{(m)}(x)I_R^{(n)}(y)\,
\theta(1-\sqrt x-\sqrt y)
\end{equation}
is the contribution to the amplitude from two chains of real 
vacuum-polarization bubbles. 
\begin{equation}
I_{R}^{(m)}(x)=-2\,{\rm Im}\,[-i\Pi(x)]^m
\label{one-chain-r}
\end{equation}
is the $m$th-order contribution to the amplitude for a single real
bubble chain with virtuality $x$.
\begin{equation}
G_{1R}=\sum_{m=1}^\infty
\int_0^1 {dx\over 2\pi x}\, f(x,0)I_R^{(m)}(x)
\label{one-gluon-total}
\end{equation}
is the amplitude for a single real bubble chain, integrated over its
virtuality and weighted with the heavy-quark and phase-space factor,
with the virtuality of the other chain set to zero.
\begin{equation}
G_{V}=\sum_{m=0}^\infty [-i\Pi(0)]^m
\label{one-chain-v}
\end{equation}
is the amplitude for a single chain of virtual vacuum-polarization
insertions. In the last term of Eq.~(\ref{total-bub}), we have used the
fact that $f(0,0)=1$.

\subsection{Manifestly Infrared-Finite Expressions}

In computing the expression for $R^{\rm Bub}$ in
Eq.~(\ref{total-bub}), it is useful to re-arrange the individual
contributions to make the expression manifestly infrared finite. First,
we define the quantity
\begin{equation}
G_1=G_{1R}+G_{V}.
\end{equation}
Since $G_{1R}$ and $G_{V}$ are the real and virtual contributions to a
single vacuum-polarization-bubble chain, the KLN theorem guarantees that
$G_1$ is infrared finite. We verify the finiteness of $G_1$ by explicit
calculation in Appendix~\ref{app:single-chain}. It is also useful to
define
\begin{eqnarray}
G_2&&=G_{2R}-G_{1R}^2\nonumber\\
&&=\sum_{m,n=1}^\infty\Biggl[\int_0^1 {dx\over 2\pi x} \int_0^1
{dy\over 2\pi y} f(x,y)I_R^{(m)}(x)I_R^{(n)}(y)\,
\theta(1-\sqrt{x}-\sqrt{y})
\nonumber\\
&&\qquad\qquad-\int_0^1 {dx\over 2\pi x} f(x,0) I_R^{(m)}(x)
\int_0^1 {dy\over 2\pi y} f(0,y) I_R^{(n)}(y)\Biggr]\,.
\label{two-gluon-finite}
\end{eqnarray}
Since the contributions from the regions of integration near $x=0$
and/or $y=0$ are equal in the first and second terms in
Eq.~(\ref{two-gluon-finite}), $G_2$ is infrared finite. Then, using
Eqs.~(\ref{total-bub}), (\ref{one-gluon-total}) and
(\ref{two-gluon-finite}), we can write
\begin{equation}
R^{\rm Bub}=R_0(G_1^2+G_2).
\label{total-bub-finite}
\end{equation}
Each term in Eq.~(\ref{total-bub-finite}) is infrared finite.

The quantity $G_1$ is computed in Appendix~\ref{app:single-chain}.
The result is
\begin{equation}
G_1={1 \over \pi \alpha_s\beta_0 }
 \arctan { { \pi \alpha_s\beta_0}\over
 {1-\alpha_s\,\beta_0 \,d}}
 +\sum_{n=1}^\infty
 \int_{0}^1 dx\, g_n(x),
\label{G1}
\end{equation}
where
\begin{equation}
g_n(x)=\lim_{\epsilon \rightarrow 0}{1\over\pi}{\rm Im}[-i\Pi(x)]^n
={1\over\pi}{\rm Im}\Bigl[\alpha_s\beta_0(d-\ln x+i\pi)\Bigr]^n,
\end{equation}
and
\begin{equation}
d=\lim_{\epsilon \rightarrow 0}{\rm Re}\, (b-a)/\epsilon.
\label{def-d}
\end{equation}
In naive non-Abelianization, 
\begin{equation}
d=\ln [\mu^2/(4m_c^2)]- C\,, 
\label{d-nna}
\end{equation}
while in the background-field gauge
\begin{equation}
d={1\over \beta_0 \pi}
\left[{67 \over 12}-{5 \over 18 } n_f-{3\over 4}(\xi^2-1)-{5\over 3}-C 
\right]+\ln {\mu^2 \over 4m_c^2}\,.
\label{d-bg}
\end{equation}

By re-arranging Eq.~(\ref{two-gluon-finite}), we can obtain an
expression for $G_2$ in which every integral is separately infrared
finite, and, hence, we can take $\epsilon$ to zero:
\begin{eqnarray}
G_2  =&& \sum_{m,n=1}^\infty\Biggl\{
-\int^1_0 {dx \over x } (1-x) g_m(x)
      \int^1_0 {dy \over y } (1-y) g_n(y)
\, \theta(\sqrt x+\sqrt y -1)\nonumber \\
 &&+\int^1_0 {dx \over x }  g_m(x)\,
       \int_0^1 {dy \over y }  g_n(y)
  [f(x,y)-(1-x)(1-y)]\,\theta(1-\sqrt x -\sqrt y)\Biggr\}.\nonumber\\
\label{G2}
\end{eqnarray}
Eqs.~(\ref{total-bub-finite}), (\ref{G1}), and (\ref{G2}) are the basis
for the calculations in this paper.

Taking into account the $\mu$ dependence of $d$
[Eq.~(\ref{def-d})], one can verify straightforwardly that
$\alpha_s(\mu) G_1 (\mu) $ and $\alpha_s^2(\mu) G_2 (\mu) $ are $\mu$
independent if the running coupling constant $\alpha_s(\mu)$ satisfies
the one-loop renormalization-group equation
\begin{equation}
(\mu \partial/\partial \mu)  \alpha_s(\mu) = 2 \beta_0 \alpha_s(\mu).
\end{equation}

\section{The Nonperturbative Regions}
\label{sec:nonpert}

So far, we have treated the effects of vacuum-polarization insertions as
if they were entirely perturbative in nature. However, we expect
perturbation theory to break down in the regions in which the gluon
virtualities $x$ and $y$ are near zero, and, for on-shell gluons, in
the regions in which energies $l_0$ and $k_0$ are near zero.

\subsection{Single Bubble Chain}
\label{sec:single-chain}

We begin by investigating the region of small virtuality for a single
bubble chain. That is, we consider the region of small $x$ in the second
term of $G_1$ in Eq.~(\ref{G1}), namely,
\begin{equation}
G_{1b}=\sum_{n=1}^\infty\int_0^1 dx\, g_n(x)
={1\over\pi }\sum_{n=1}^\infty\int_0^1 dx\, {\rm Im}
\Big[\alpha_s\beta_0(d-\ln x+i\pi)\Big]^n.
\label{G1b}
\end{equation}
Near $x=0$, the $\ln x$ in $G_{1b}$ becomes unbounded, and the series
fails to converge. This means that, in principle, the result that one
obtains by carrying out the integration before the summation may be
different from the result that one obtains with the opposite order of
operations. In fact, as we shall see, interchange of these operations
does produce a finite difference in the result.

Since
\begin{equation}
\int_0^\lambda dx\,\ln ^n x \sim n!,
\end{equation}
one might expect the terms in the series in Eq.~(\ref{G1b}) to exhibit
factorial growth, arising from the region of integration near $x=0$.
In fact, a careful calculation shows that the factorial growth cancels
when one takes the imaginary part. Factorial growth in a bubble-chain
series is associated with a renormalon singularity in the Borel
transform of the series. Explicit calculation of the Borel transform of
$G_{1b}$ shows that the renormalon singularity in the Borel plane
vanishes when one takes the imaginary part \cite{beneke}.

It is useful, in understanding both the dependence on the order of
operations and the absence of factorial growth, to re-write the
expression for $G_{1b}$ as a contour integral:
\begin{equation}
G_{1b}={1\over 2\pi i}\sum_{n=1}^\infty\int_C dz\,
\Big\{\alpha_s\beta_0[d-\ln (-z)]\Big\}^n.
\label{G1b-contour}
\end{equation}
The contour $C$ runs from $z=1-i\epsilon$ to $z=0$ to $z=1+i\epsilon$.
Term by term in the series, we can, without encountering any
singularities, deform the contour $C$ into the contour $C'$, which is a
circle of unit radius, centered at the origin, and traversed in the 
clockwise direction from $z=1-i\epsilon$ to $z=1+i\epsilon$. Along the
new contour $C'$, the magnitude of $z$ is never small. Hence, the
logarithm in Eq.~(\ref{G1b-contour}) is bounded, and it is clear that
there is no factorial growth in the series. Furthermore, the
perturbation sum is now uniformly convergent, and we can interchange the
summation and integration without affecting the result. Hence,
\begin{equation}
G_{1b}={1\over 2\pi i}\int_{C'} dz\, {1\over 1-\alpha_s\beta_0
[d-\ln (-z)]}\,.
\label{G1b-cp}
\end{equation}

Suppose, on the other hand, that we carry out the perturbation summation
before deforming the contour. Then we have
\begin{equation}
G'_{1b}={1\over \pi}\int_0^1 dx\, {\rm Im}\left[
{1\over 1-\alpha_s\beta_0(d-\ln x +i\pi)}\right]
={1\over 2\pi i}\int_C dz\, {1\over 1-\alpha_s\beta_0
[d-\ln (-z)]}\,.
\label{Gp1b}
\end{equation}
If we deform the contour to $C'$, then we encounter the Landau pole at
\begin{equation}
z_0=-\exp\left[d-{1\over \alpha_s\beta_0}\right],
\label{landau-pole}
\end{equation}
and we must include its residue in the result:
\begin{equation}
G'_{1b}={1\over 2\pi i}\int_{C'} dz\, {1\over 1-\alpha_s\beta_0
[d-\ln (-z)]}+{z_0\over \alpha_s\beta_0}.
\label{Gp1b-cp}
\end{equation}
Then, using Eqs.~(\ref{G1b-cp}) and (\ref{Gp1b-cp}), we have
\begin{equation}
G_{1b}=G'_{1b}-{z_0\over \alpha_s\beta_0}.
\label{G-Gp}
\end{equation}
The contribution from the residue at the Landau pole [the second term in
Eq.~(\ref{G-Gp})] is just the difference in results between carrying
out the integration before the summation and carrying out the summation
before the integration. It displays an essential singularity in
$\alpha_s$ and has the characteristic form of a power correction
associated with a renormalon singularity. However, as we have seen, in
this case, there is no renormalon singularity associated with the Borel
transform of the series.

The ambiguity in the result for $G_{1b}$ raises a question as to which
order of operations, if either, is correct. If we carry out the
integration before the summation, then the contour-deformation argument
shows that there is no contribution from the region of small virtuality.
In this case, asymptotic freedom suggests that the perturbation series
is reliable. On the other hand, if we carry out the summation first,
then there is no reason to trust the resulting expression: for fixed $x$
sufficiently small, we are outside the radius of convergence of the
geometric series and outside the regime of asymptotic freedom. Indeed,
the Landau pole in the summed series is most likely an artifact of
perturbation theory that would not persist in the true solution of QCD
and, hence, the low-virtuality contribution from the residue at the pole
is spurious.\footnote{Note that the situation is less clear in the
present case than, for example, in the analysis of the $\tau$ hadronic
width, for which similar contour deformation arguments have been made
\cite{braaten-narison-pich}. The $\tau$ width possesses a spectral
(K\"allen-Lehmann) representation, which gives one some information
about its analytic properties. The gluon propagator, being a colored
object, possesses no spectral representation.} Therefore, we conclude
that the correct order of operations is to perform the integration first
and then the summation. 

We note that the procedure of integrating over the gluon virtualities
before carrying out the perturbation summation is equivalent to the
standard Borel-transform method \cite{beneke} for evaluating the
contribution of a bubble chain. In that method, one takes the Borel
transform of the series, sums the transformed series, carries out the
integration over the phase space, and, finally, takes the inverse Borel
transform. Since the initial Borel transform renders the series
convergent for all virtualities and absolutely integrable, one can
interchange the order of operations, carrying out the integration first
and then summing the Borel-transformed series. Furthermore, term by term
in the series, the integration and Borel-transformation commute.
Therefore, the standard Borel-transform method is equivalent to carrying
out the integration, taking the Borel transform, summing the series, and
then taking the inverse transform. This, by the definition of the
inverse transform, is equivalent to carrying out the integration and
then summing the perturbation series.

In Appendix~\ref{app:single-chain}, we have computed the first term of
$G_1$ in Eq.~(\ref{G1}) by performing the integration over $x$ before
the summation of the perturbation series. In this case, that is the only
sensible order of operations: the perturbation series is not convergent
for small values of the dimensional regulator $\epsilon$ until one has
canceled the infrared divergences by carrying out the integration over
$x$ in the real contribution and adding the real and virtual
contributions order by order.

We must also consider the possibility that, when $x$ ($y$) is
sufficiently close to unity, the energy $l_0$ ($k_0$) of the on-shell
gluon may be too small to lie within the perturbative regime. (Recall
that, although we have written Eq.~(\ref{G1b}), for compactness, only
in terms of $x$, it derives from expressions that are symmetric in $x$
and $y$.) Let us, for the moment, set aside the contributions from the
nonperturbative regions of small $l_0$ ($k_0$) by imposing a restriction
on the region of integration:
\begin{equation}
x,y<(1-\sqrt\delta)^2.
\label{delta}
\end{equation}
Here, $\delta$ is a positive parameter that we take to be much less than
unity, but not so small that $4m_c^2\delta$ is outside the perturbative
regime. Using Eq.~(\ref{kinematics}), we see that the restriction
(\ref{delta}) implies that
\begin{mathletters}
\begin{eqnarray}
l_0^2&&\geq m_c^2(2\sqrt \delta-\delta)^2\approx 4m_c^2\delta,\\
k_0^2&&\geq m_c^2(2\sqrt \delta-\delta)^2\approx 4m_c^2\delta.
\end{eqnarray}
\label{l0-k0}%
\end{mathletters}
As we shall see in Sec.~\ref{sec:matrix-element}, the nonperturbative
regions that we have set aside in imposing this restriction yield
contributions of the form of one-loop corrections to the NRQCD matrix
element of a color-octet 4-fermion operator.

\subsection{Two Bubble Chains}
\label{sec:2-chains}

The quantity $G_2$ [Eq.~(\ref{G2})] arises the from real-real
contribution of the bubble chains. Hence, both virtualities $x$ and $y$
are nonzero. Furthermore, there are constraints on $x$ and $y$. The
first term in braces is subject to the constraint $\sqrt x +\sqrt y\geq
1$; the second term in braces is subject to the constraint $\sqrt x
+\sqrt y\leq 1$. Owing to these constraints, one cannot immediately
apply the contour-deformation argument of the last subsection to avoid
the regions of small virtuality. For the first term in braces, when $x$
is near unity, $y$ ranges from unity almost to zero [$(1-\sqrt x)^2$].
One could write the integral over $y$ as a contour integral in the
complex plane. However, the lower endpoints are tied down at $y=(1-\sqrt
x)^2\pm i\epsilon$, and, hence, one cannot deform the contour out of the
small-$y$ region. For the second term in braces, when $x$ is near unity,
then $y$ ranges from zero up to a small number [$(1-\sqrt x)^2$]. Again,
one could write the integral over $y$ as a contour integral in the
complex plane, but owing to the fixed endpoints at $y=(1-\sqrt x)^2\pm
i\epsilon$, one cannot deform the contour out of the small-$y$ region.
We deal with this difficulty by setting aside, temporarily, parts of the
region of integration by imposing, again, the restriction (\ref{delta}).

For the first term in braces in Eq.~(\ref{G2}), the restriction
(\ref{delta}), combined with the constraint $\sqrt x +\sqrt y\geq 1$,
guarantees that
\begin{equation}
x,y\geq\delta.
\end{equation}
Hence, the $x$ and $y$ integrations lie entirely within the perturbative
regime.

For the second term in braces in Eq.~(\ref{G2}), we may use the
restriction (\ref{delta}) to decompose the region  of integration as
follows:
\begin{eqnarray}
&&\int_0^{(1-\sqrt \delta)^2} dx\,
\int_0^{(1-\sqrt \delta)^2}dy\,\theta(1-\sqrt x-\sqrt y)
=\int_\delta^{(1-\sqrt \delta)^2}dx\,
\int_\delta^{(1-\sqrt \delta)^2}dy\;\theta(1-\sqrt x-\sqrt y)\nonumber\\
&&\qquad+\int_\delta^{(1-\sqrt \delta)^2}dx\,\int_0^\delta dy\,
+\int_0^\delta dx\,\int_\delta^{(1-\sqrt \delta)^2}dy\,
+\int_0^\delta dx\, \int_0^\delta dy\,.
\label{region}
\end{eqnarray}
In the first term on the right side of Eq.~(\ref{region}), both
integrations lie entirely within the perturbative regime. In the second
term on the right side of Eq.~(\ref{region}), the $x$ integration lies
entirely within the perturbative regime, while the $y$ integration does
not. However, we can write the $y$ integral, as in the preceding
subsection, as an integral over a contour running from
$\delta-i\epsilon$ to zero to $\delta+i\epsilon$. We can then deform
this contour into a circle of radius $\delta$ centered at the origin
that is traversed in the clockwise direction from $\delta-i\epsilon$ to
$\delta+i\epsilon$. It is easy to see that one encounters no
singularities in the integrand in performing this contour deformation.
Hence, we conclude that there are no contributions to the second term on
the right side of Eq.~(\ref{region}) from the region of small $y$.
Similarly, for the third term on the right side of Eq.~(\ref{region}),
we can write that the $x$ integral as a contour integral and deform
the contour out of the region of small $x$. Finally, for the fourth term
on the right side of Eq.~(\ref{region}), we can write both the $x$ and
$y$ integrals as contour integrals and deform both contours out of
the region of small virtuality. Therefore, we conclude that, once we
have imposed the restriction (\ref{delta}), the contributions to the
second term in braces in Eq.~(\ref{G2}) lie entirely within the
perturbative regime.

Again, we shall find in Sec.~\ref{sec:matrix-element} that the
contributions that we have set aside by virtue of the restriction
(\ref{delta}) correspond to one-loop corrections to the NRQCD matrix
element of a color-octet 4-fermion operator. We note that the
contributions to $G_2$ that we have set aside, when treated in
perturbation theory, yield renormalon singularities. Such singularities
are a signal that the contribution cannot be computed reliably in
perturbation theory. In Appendix~\ref{app:renormalon}, we compute the
leading renormalon singularity that arises from these contributions to
$G_2$.

\subsection{Contribution of the Color-Octet NRQCD Matrix Element}
\label{sec:matrix-element}%

Now let us discuss the nonperturbative contributions that we have set
aside by virtue of the restriction on the region  of integration
(\ref{delta}). For the contributions that have been set aside, we have
either $x\geq (1-\sqrt{\delta})^2$ or $y\geq(1-\sqrt{\delta})^2$, but not
both, since this would violate the kinematic limit
$\sqrt{x}+\sqrt{y}<1$. Hence, one of the final-state bubble chains is
highly virtual compared to the momentum of the other bubble chain or the
relative momentum of the incoming heavy $Q\bar Q$ pair. Therefore, we
may approximate these contributions by shrinking the highly virtual
bubble chain to a point, replacing the bubble chain and its $Q\bar Q$
interaction vertices with a four-fermion interaction. Corrections to this
approximation may be taken into account by including four-fermion
interactions containing derivatives with respect to the heavy-quark
separations. It follows that the contributions that have been set aside
are actually one-bubble-chain corrections to the matrix element of a
four-fermion operator in the $\eta_c$ state.

In NRQCD at leading order in $v$, the four-fermion operator that arises
when one shrinks a bubble chain to a point is the color-octet,
spin-triplet operator
\begin{equation}
{\cal O}_8(^3S_1)=\psi^+ \sigma^i T^a \chi \chi^\dagger \sigma^i T^a
\psi\,.
\label{octet-op}
\end{equation}
We now verify that the one-bubble-chain corrections to the contributions
of this operator to $R$ do indeed reproduce, at leading order in $v$, the
contributions that we have set aside.

According to the NRQCD factorization formula \cite{bbl} the
contribution of this operator to the decay rate factors:
\begin{equation}
\Gamma_8^{\rm Bub}(\eta_c)=2\,{\rm Im} \,
\left[{f_8(^3S_1)\over m_c^2}\right]
\langle\eta_c|{\cal O}_8(^3S_1)|\eta_c\rangle \,,
\label{octet}
\end{equation}
where ${\rm Im}\,[f_8(^3S_1)/m_c^2]$ is the short-distance coefficient.
By matching Eq.~(\ref{octet}) to the imaginary part of the amplitude in full
QCD for $Q\bar Q\rightarrow Q\bar Q$ via a single bubble chain, one
finds, in the bubble-chain approximation, that
\begin{equation}
2\,{\rm Im }\, \left[{f_8(^3S_1)\over m_c^2}\right]=
-2g^2\,{\rm Im} \,[K(1)]\,,
\label{short-distance-octet}
\end{equation}
where $K(x)$ is defined in Eq.~(\ref{bub-prop}).

The color-octet matrix element is fully determined by the NRQCD effective
Lagrangian. It is nonperturbative in nature and must be determined
phenomenologically or through methods such as lattice QCD. However, our
aim here is to compute the perturbative contribution to this matrix
element in the bubble-chain approximation so that we can compare it with
the part of our perturbative calculation of $R^{\rm Bub}$ that we have
set aside.

In NRQCD, the one-bubble-chain correction to the operator matrix element
arises, in leading order in the heavy-quark velocity $v$, through the
insertion of $\sigma \cdot B$ operators into any two of the four quark
lines. In the Feynman gauge, this contribution can be written as
\begin{eqnarray}
\langle H| {\cal O}_8(^3S_1)|H\rangle
&&=  4g^2 {C_F \over 2 N_c} \sum_{m=0}^\infty
\int^\Lambda { d^4 k \over
(2\pi)^4 }\, i K^{(m)}(x)\,{({\mathbf{\sigma}} \times
{\mathbf{k}} )_i \over 2m_c }  {(-{\mathbf{\sigma}} \times
{\mathbf{k}} )_i \over 2m_c } \nonumber\\
&&\qquad\times \left({i \over -k_0 -{{\mathbf
k}^2 /(2m_c)} +i \epsilon }\right)^2 \,
\langle H|{\mathcal{O}}_1({}^1S_0)|H\rangle \nonumber
\\
&&=
  -4g^2{C_F \over 2 N_c} \sum_{m=0}^\infty
\int^\Lambda { d^4
k \over (2\pi)^4 }\, iK^{(m)}(x)\left({1 \over -k_0
-{{\mathbf k}^2/(2m_c)}+i \epsilon }\right)^2
{{\mathbf{k}}^2 \over 2m_c^2}\;\nonumber\\
&&\qquad\times\langle H|{\mathcal{O}}_1(^1S_0)|H\rangle. \label{M-E}
\end{eqnarray}
The superscript $\Lambda$ is a reminder that the NRQCD effective theory
requires an ultraviolet cutoff. $H$ denotes a hadronic state.

The $k_0$ integral can be re-written by making use of the techniques of
contour integration. Closing the $k_0$ contour in the lower half plane,
we pick up the residues at the poles and the discontinuities across the
cuts in $K^{(m)}(x)$. The contribution of these residues and
discontinuities can be written in terms of the imaginary part of
$K^{(m)}(x)$. Then we have
\begin{eqnarray}
\langle H|{\cal O}_8({}^3S_1)|H\rangle &&=
   {g^2 C_F \over  N_c m_c^2} \sum_{m=0}^\infty
\int^\Lambda { d^4
k \over (2\pi)^4 }\, \theta(k_0)[-2\,{\rm Im} K^{(m)}(x)]
\left({1 \over -k_0 -{{\mathbf k}^2 /(2m_c)} +i
\epsilon }\right)^2 {{\mathbf{k}}^2 } \;\nonumber\\
&&\qquad\times\langle H|{\mathcal{O}}_1({}^1S_0)|H\rangle.
\label{k0-int}
\end{eqnarray}

Using Eq.~(\ref{kinematics}), we can replace integration variables $k_0$
and $\mathbf{k}$ with $x$ and $y$. We note that
\begin{equation}\label{dxdy}
  d k_0 \, d |{\mathbf{k}}|^2 = 4 m_c^3\, dx \,dy.
\end{equation}
Making this change of variables in Eq.~(\ref{k0-int}) and multiplying by
the short-distance quantity in Eq.~(\ref{short-distance-octet}), we
obtain the one-bubble-chain correction to the rate:
\begin{equation}
\Gamma_8^{\rm Bub}(H)=
{2\pi\alpha_s^2 C_F \over N_c m_c^2} \sum_{m=0}^\infty
\int_0^\Lambda {dx\over 2\pi}\,
2\,{\rm Im}\left[4m_c^2 K(1)\right] \,2\,{\rm Im}
\left[4m_c^2 K^{(m)}(x)\right] f(x,y)\, \langle H|{\cal O}_1(^1S_0)|H\rangle,
\label{octet-rate}
\end{equation}
where $f(x,y)$ is defined in Eq.~(\ref{f-xy}), and we have used the
approximate relation
\begin{equation}\label{l0}
k_0 + {{\mathbf{k}}^2 \over 2m_c } \approx k_0 - {k^2 \over
2m_c}  \,,
\end{equation}
which holds for $k_0/m_c<<1$. Dividing the rate in Eq.~(\ref{octet-rate})
by the rate into two photons in Eq.~(\ref{2-photons}), we obtain the
contribution of the color-octet matrix element to $R^{\rm Bub}$:
\begin{equation}
R_8^{\rm Bub}=R_0\sum_{m=0}^\infty \int_0^\Lambda {dx\over 2\pi}\,
{dy\over 2\pi}\,
2\,{\rm Im} \left[4m_c^2 K(1)\right] \,2\,{\rm Im}
\left[4m_c^2 K^{(m)}(y)\right]f(x,y).
\label{R-octet}
\end{equation}

Now we choose the cutoff $\Lambda$ on the matrix element to be the hard 
cutoff
\begin{eqnarray}
x&&\leq\delta;\nonumber\\
y&&\geq (1-\sqrt\delta)^2.
\end{eqnarray}
From Eq.~(\ref{kinematics}), we see that this choice of cutoff
corresponds to
\begin{eqnarray}
k^2&&\leq 4m_c^2\delta;\nonumber\\
{\bf k}^2&&\leq [k^2/(4m_c)+2m_c\sqrt\delta-m_c\delta]^2-k^2.
\label{mom-cutoff}
\end{eqnarray}
Then, symmetrizing Eq.~(\ref{R-octet}) in $x$ and $y$, we find that it
is precisely the contribution to Eq.~(\ref{R-bub}) that is excluded
by the restriction (\ref{delta}). 

We conclude that the contributions that we have set aside by virtue of
the restriction (\ref{delta}) correspond to contributions to the decay
rate that are proportional to the matrix element of the color-octet
operator ${\cal O}_8({}^3S_1)$. In deriving Eq.~(\ref{R-octet}), we have
dropped contributions of higher order in $v$. These could be taken into
account by including the contributions of higher-dimensional color-octet
operators.

We note that the use of a hard cutoff in the color-octet matrix element,
as opposed to dimensional regularization, is essential in order to
contain the small-virtuality contributions entirely within the matrix
element \cite{bodwin-chen}. These contributions have an
ultraviolet-divergent power behavior. In dimensional regularization,
power-divergent contributions are set to zero, and, therefore, the
small-virtuality contributions would be excluded from a
dimensionally-regulated matrix element.

Using Eqs.~(\ref{octet}) and (\ref{short-distance-octet}), along with
Eqs.~(\ref{r0}) and (\ref{2-photons}), we see that the contribution of
the color-octet operator ${\cal O}_8({}^3S_1)$ to $R$ is given by
\begin{equation}
R_8=-2\,{\rm Im}\, [4m_c^2K(1)]R_0{N_c\over \alpha_sC_F}
{\langle\eta_c|{\cal O}_8({}^3S_1)|\eta_c\rangle\over 
\langle\eta_c|{\cal O}_1({}^1S_0)|\eta_c\rangle}\,.
\label{octet-nonpert}
\end{equation}
We note that, in the limit in which the dimensional regulator $\epsilon$ 
is taken to zero, 
\begin{equation}
{\rm Im}\, [4m_c^2K(1)]={\alpha_s\beta_0\pi\over 
(1-\alpha_s\beta_0 d)^2+\pi^2\alpha_s^2\beta_0^2}\,.
\label{K1}
\end{equation}

In this paper, we will not compute $R_8$, but, instead, will treat it as
an uncertainty in the calculation. The ratio of matrix elements in
Eq.~(\ref{octet-nonpert}) may be estimated by making use of the
velocity-scaling rules for NRQCD \cite{bbl}. If we adopt the suggestion
of Petrelli {\it et al.} \cite{Petrelli:1998ge}, and include, in
addition to the factors of the velocity $v$, a factor $1/(2N_c)$ in the
estimate of the color-singlet matrix element, then we obtain
\begin{equation}
{\langle\eta_c|{\cal O}_8({}^3S_1)|\eta_c\rangle\over
\langle\eta_c|{\cal O}_1({}^1S_0)|\eta_c\rangle}\sim {v^3\over 2N_c}.
\label{ratio-vel}
\end{equation}
A one-loop perturbative estimate gives a result that is numerically very
close to that in Eq.~(\ref{ratio-vel}):
\begin{equation}
{\langle\eta_c|{\cal O}_8({}^3S_1)|\eta_c\rangle\over
\langle\eta_c|{\cal O}_1({}^1S_0)|\eta_c\rangle}
\sim {v^3C_F\over \pi N_c}.
\label{ratio-one-loop}
\end{equation}
In making numerical estimates of the size of $R_8$, we will use the
expression (\ref{ratio-one-loop}). 

The value of the color-octet matrix element depends on the cutoff
$\delta$. In making the estimates (\ref{ratio-vel}) and
(\ref{ratio-one-loop}), one assumes that the cutoff, in terms of
momentum, is of the order of the dynamical scale $mv$. On the other
hand, we would like to choose $\delta$ so that nonperturbative
contributions arising from momenta less than $v$ are contained in the
color-octet operator matrix element, while contributions arising from
larger momenta are contained in the short-distance coefficients. We will
assume that a momentum cutoff of order $mv$ is consistent with these
desiderata. From Eq.~(\ref{mom-cutoff}), we see that 
\begin{equation}
{\bf k}^2/m_c^2\leq(2\sqrt \delta-\delta)^2.
\end{equation}
The choice $\delta=0.1$ yields ${\bf k}^2/m_c^2\leq 0.28.$ For
charmonium, $0.28$ is very close to the average value of $v^2$.
Therefore, we will use the value $\delta=0.1$ in our calculations.

\section{Efficient Method for Numerical Computation}
\label{sec:efficient}

The perturbation series in Eq.~(\ref{G1}) is convergent. Once we have
imposed the restriction (\ref{delta}), the series in the first and
second terms of Eq.~(\ref{G2}) are also convergent. However, for
reasonable choices of $\delta$, the series converge very slowly. For
example, for the second term in Eq.~(\ref{G2}) with $\delta=0.1$ and
$\alpha_s=0.247$, it is necessary to compute through 12th order in
$\alpha_s$ in order to achieve an accuracy of about 3\%. Furthermore, at
high orders in $\alpha_s$, the integrand becomes strongly oscillatory,
and the integral is difficult to compute numerically. It would be far
more efficient if we could carry out the perturbation summation before
integrating over $x$ and $y$.

We have already seen in Sec.~\ref{sec:single-chain} that, in the case of
$G_1$, we can carry out the perturbation sum first, provided that we
compensate by adding a contribution that is proportional to the residue
at the Landau pole. Using Eqs.~(\ref{G1b}), (\ref{Gp1b}) and (\ref{G-Gp}), 
we can write 
\begin{equation}
\sum_{n=1}^\infty \int_0^1 dx\, g_n(x)=\int_0^1 dx\, g(x)
-{z_0\over \alpha_s\beta_0},
\end{equation}
where
\begin{equation}
g(x)\;=\sum_{n=1}^\infty g_n(x)=
{\alpha_s\beta_0 \over \left[1 - \alpha_s\beta_0 (d -\ln x)
\right]^2  + (\alpha_s\beta_0 \pi)^2  } \,.
\end{equation}
Then, from Eq.~(\ref{G1}), we have
\begin{equation}
G_1(\delta)={1 \over \pi \alpha_s\beta_0 }
 \arctan { { \pi \alpha_s\beta_0}\over
 {1-\alpha_s\,\beta_0 \,d}}
+\int_{(1-\sqrt\delta)^2}^1 dx\, {g(x)\over x}
+ \int_{0}^{(1-\sqrt\delta)^2} dx\, g(x)
-{z_0\over \alpha_s\beta_0}.
\label{G1-mod}
\end{equation}
The second term in Eq.~(\ref{G1-mod}) arises from applying the cutoff 
(\ref{delta}) to $G_{1Rb}$ [Eq.~(\ref{bub:Rb})].

The analysis of Eq.~(\ref{G2}) is somewhat more involved. For the first 
term in Eq.~(\ref{G2}), the restriction on the region  of integration
(\ref{delta}), together with the constraint $\sqrt x+\sqrt y\leq 1$,
guarantees that $x,y\geq \delta$. Then the sums over $n$ and $m$ are 
absolutely convergent, provided that $\delta$ is sufficiently large that 
\begin{equation}
|d-\ln\delta+i\pi|\alpha_s\beta_0<1.
\label{delta-max}
\end{equation}
For $\alpha_s=0.247$, Eq.~(\ref{delta-max}) requires that $\delta$ be
greater than 0.0482 for the NNA value of $d$ [Eq.~(\ref{d-nna})] and
greater than 0.0752 for the background-field-gauge value of $d$ with
$\xi=1$ [Eq.~(\ref{d-bg})]. Then, the convergence of the series and the
absolute integrability of the sums allow us to interchange the summations
and the integrations.

For the second term in Eq.~(\ref{G2}), we decompose the region of
integration according to Eq.~(\ref{region}). For the first term on the
right side of Eq.~(\ref{region}), the perturbation series are absolutely
convergent, provided that Eq.~(\ref{delta-max}) is satisfied, and we can
interchange the summations and the integrations. For the second term on
the right side of Eq.~(\ref{region}), the $x$ integration lies within the
region of absolute convergence of the series, but the $y$ integration
does not. However, one can apply the contour-deformation analysis of
Sec.~\ref{sec:single-chain} to the $y$ integral, deforming the contour
to a circle of radius $\delta$ centered at the origin, so that the
entire contour lies within the region of absolute convergence. [The
additional factors in the integrand of the second term in Eq.~(\ref{G2})
contain no singularities within that circle.] It follows that we can
interchange the $y$ integration with the summations, provided that we
add a compensating term that is proportional to the residue at the
Landau pole. Similarly, for the third term on the right side of
Eq.~(\ref{region}), we can interchange the integrations with the
summations, provided that we add a compensating term to the $x$ integral
that is proportional to the residue at the Landau pole. For the fourth
term on the right side of Eq.~(\ref{region}), we must add
Landau-pole-residue terms to both the $x$ and $y$ integrals in order to
interchange the integrations with the summations. Then, the various
terms can be recombined to obtain
\begin{eqnarray}
G_2(\delta) =&& 
-\int^{(1-\sqrt\delta)^2}_0 {dx \over x } (1-x) g(x)
      \int^{(1-\sqrt\delta)^2}_0 {dy \over y } (1-y) g(y)
\, \theta(\sqrt x+\sqrt y -1)\nonumber \\
&&+\int_0^{(1-\sqrt\delta)^2}{dx\over x}g(x)
\int_0^{(1-\sqrt\delta)^2}{dy\over y}g(y)\,
[f(x,y)-(1-x)(1-y)]\,
\theta(1-\sqrt x -\sqrt y)\nonumber\\
&&-{2z_0\over \alpha_s\beta_0}\int_0^{(1-\sqrt\delta)^2}
{dx\over x}g(x)\,[f(x,z_0)-(1-x)(1-z_0)]\nonumber\\
&&+{z_0^2\over \alpha_s^2\beta_0^2}\,[f(z_0,z_0)-(1-z_0)^2].
\label{G2-mod}
\end{eqnarray}

We have verified that the expressions (\ref{G1-mod}) and (\ref{G2-mod})
agree, within truncation errors, with numerical evaluations of the
perturbation sums for $G_1$ and $G_2$ through 12th order in $\alpha_s$.

\section{Results}
\label{sec:results}

In order to combine our result from the resummation of
vacuum-polarization bubbles with the computation through order
$\alpha_s$, we must subtract the part of the contribution through
relative order $\alpha_s$ that is contained in the resummation.
Expanding $G_1$ [Eq.~(\ref{G1})] through order $\alpha_s$, we obtain
\begin{equation}
G_1=1+\alpha_s\beta_0(1+d)+\ldots.
\end{equation}
The perturbation series for $G_2$ [Eq.~(\ref{G2})] begins at order
$\alpha_s^2$. Therefore, using Eq.~(\ref{total-bub-finite}), we see that
we must subtract $R_0[1+2\alpha_s\beta_0(1+d)]$. We conclude that the
rate $R$, including the exact next-to-leading-order contribution and
bubble resummation of the higher-order contributions, is given by
\begin{eqnarray}
R^{\rm Res}=R_0(\mu)\Biggl\{&&G_1^2(\mu)+G_2(\mu)
+\left({199\over 6}-{13\pi^2\over 8}
-{8\over 9}n_f\right){\alpha_s(\mu)\over\pi}\nonumber\\
&&+2\alpha_s(\mu)\beta_0\Biggl[\ln{\mu^2\over 4 m_c^2}
-1-d(\mu)\Biggr]\Biggr\}.
\end{eqnarray}
For the case of NNA resummation, we have 
\begin{equation}
R^{\rm NNA}=R_0(\mu)\left\{\Bigl[G_1^{\rm NNA}(\mu)\Bigr]^2
+G_2^{\rm NNA}(\mu)
+\left({37\over 2}-{13 \pi^2\over 8}\right)
{\alpha_s(\mu) \over \pi }\right\}\,,
\label{r:resum}
\end{equation}
while for the case of background-field-gauge resummation, we have
\begin{eqnarray}
R^{\rm BFG}=R_0(\mu)
\Biggl\{&&\Bigl[G_1^{\rm BFG}(\mu)\Bigr]^2+G_2^{\rm BFG}(\mu)
+\Biggl[22-{13 \pi^2\over 
8}-{n_f\over 3}-{3\over 2}(\xi^2-1)\Biggr]
{\alpha(\mu)\over\pi}\nonumber\\
&&-2\alpha_s(\mu)\beta_0
{\alpha_s(\mu) \over \pi }\Biggr\}\,,
\end{eqnarray}
where $\xi$ is the gauge parameter for fields internal to loops.
($\xi=1$ in the Feynman gauge.)

We now evaluate $R^{\rm Res}$, taking $n_f=3$, choosing $\mu=2m_c$, with
$m_c$ the charm-quark pole mass, and using $\alpha_s^{(3)}(2m_c)=0.247\pm
0.012$.\footnote{We have obtained this value by evolving with 3-loop
accuracy from $\alpha_s(m_Z)=0.118\pm 0.002$, where $m_Z=91.188~{\rm
GeV}$ is the $Z$-boson mass, and using for the $\overline {\rm MS}$
charm-quark and bottom-quark masses $m_c^{\overline {\rm MS}}=1.25\pm
0.10~{\rm GeV}$ and $m_b^{\overline {\rm MS}}=4.2\pm 0.2~{\rm GeV}$,
respectively \cite{PDG}. The uncertainty in $\alpha_s(2m_c)$ includes
the uncertainty in $\alpha_s(M_Z)$ and the uncertainty in the value of
$m_c$.} We use Eqs.~(\ref{G1-mod}) and (\ref{G2-mod}) to compute $G_1$
and $G_2$, respectively, with the cutoff $\delta=0.1$.

In the case of the NNA resummation, we obtain 
\begin{eqnarray}
&&G_1^{\rm NNA}=1.67,\nonumber\\
&&G_2^{\rm NNA}=-0.64,\nonumber\\
&&R^{\rm NNA}=(3.01\pm 0.30\pm 0.34)\times 10^3,
\end{eqnarray}
where the first uncertainty in $R^{\rm NNA}$ comes from the uncertainty
in $\alpha_s(2m_c)$, and the second uncertainty comes from the estimate
of the uncalculated color-octet contribution [Eqs.~(\ref{octet-nonpert}),
(\ref{K1}), and (\ref{ratio-one-loop})], with $v=0.3$.

In the case of background-field-gauge resummation, we take the gauge
parameter $\xi$ to be equal to unity, since this choice minimizes the
size of the residual relative-order-$\alpha_s$ contribution. Then we
obtain
\begin{eqnarray}
&&G_1^{\rm BFG}=1.88,\nonumber\\
&&G_2^{\rm BFG}=-1.04,\nonumber\\
&&R^{\rm BFG}=(3.26\pm 0.31\pm 0.47)\times 10^3, 
\end{eqnarray}
where the uncertainties in $R^{\rm BFG}$ are as in $R^{\rm NNA}$.

There are additional uncertainties in our results that arise from the
dependences on the renormalization scale $\mu$. We expect the scale
dependences in the resummed expressions to be considerably less than in
the NLO expression. As we have already mentioned,
$R_0(\mu)[G_1^2(\mu)+G_2(\mu)]$ is invariant with respect to changes of
scale at the level of one-loop running of $\alpha_s(\mu)$. The
remaining, unresummed terms in $R^{\rm Res}$ are a small fraction of the
total expression and yield a correspondingly small dependence on the
scale. If we evolve from $\alpha_s(2m_c)=0.247$ using the one-loop beta
function, then we obtain $\alpha_s(m_c)=0.327$. Taking this value of
$\alpha_s(m_c)$ and $\mu=m_c$, we obtain a 10.6\% increase in the value
of $R^{\rm NNA}$ and a 1.8\% increase in the value of $R^{\rm BFG}$. At
this value of $\alpha_s(m_c)$, the inequality (\ref{delta-max}) is no
longer satisfied for the background-field-gauge resummation, and the
perturbation expansion no longer converges. For purposes of estimating
the scale dependence, we assume that $G_1$ and $G_2$ are given by
Eqs.~(\ref{G1-mod}) and (\ref{G2-mod}) and, therefore, that
$R_0(\mu)[G_1^2(\mu)+G_2(\mu)]$ is scale invariant.

Under three-loop evolution of $\alpha_s(\mu)$,
$R_0(\mu)[G_1^2(\mu)+G_2(\mu)]$ is no longer scale invariant, and,
hence, the scale dependence is larger. Three-loop evolution from
$\alpha_s(2m_c)=0.247$ yields $\alpha_s(m_c)=0.350$. At this value of
$\alpha_s(m_c)$, the inequality (\ref{delta-max}) is not satisfied in
either the NNA or background-field-gauge resummations, and we again
assume, for purposes of estimating the scale dependence, that $G_1$ and
$G_2$ are given by Eqs.~(\ref{G1-mod}) and (\ref{G2-mod}). Taking the
three-loop-evolved value of $\alpha_s(m_c)$ and $\mu=m_c$, we obtain a
24.5\% increase in the value of $R^{\rm NNA}$ and an 11.9\% increase in
the value of $R^{\rm BFG}$. One could hope to reduce these uncertainties
further through a resummation of two-loop bubble contributions. Still,
they are already a significant improvement over the factor of 2.4
uncertainty from scale dependence in the NLO result.

\section{Summary and Discussion}
\label{sec:discussion}

We have studied the ratio $R$ of the hadronic and electronic decay rates
for the heavy quarkonium $\eta_c$. In $R$, the dependences of the rates
on NRQCD matrix elements cancel at leading order in $v$. The
relative-order $v^2$ corrections also cancel. Consequently, renormalon
ambiguities associated with initial-state virtual gluons cancel at this
level of accuracy in $v^2$ \cite{braaten-chen,bodwin-chen}.

The perturbative corrections to $R$ that are associated with the running
of $\alpha_s$ account for most of the large one-loop correction to $R$.
We have carried out a resummation of the leading corrections of this
type, which arise from vacuum-polarization insertions in the final-state
gluon legs. Our results are in good agreement with the experimental
value of $R$.

There is some arbitrariness in the choice of the resummed quantity. In
this paper, we have used the NNA approach \cite{naive-non}, in which one
resums, in $n$th order, all contributions proportional to
$(\alpha_s\beta_0)^n$. In implementing this method, one resums chains of
quark-loop vacuum-polarization bubbles in gluon legs and then accounts
for gluon contributions by promoting the quark contribution to $\beta_0$
to the full QCD $\beta_0$. An alternative to the NNA approach is to
resum chains of quark-loop {\it and} gluon-loop vacuum-polarization
bubbles in the background-field gauge \cite{background-field}. This
method is motivated by the fact that, in the background-field gauge, all
of the running of $\alpha_s$ is contained in the vacuum-polarization
diagrams \cite{abbott}. The two approaches give numerical results that
agree at the level of the other uncertainties in the calculations.

The sum of the vacuum-polarization-bubble terms is independent of the
renormalization-scale, at the level of the one-loop running of
$\alpha_s$. Hence, the renormalization-scale dependence of our result is
much milder than that of the unresummed one-loop result. However,
because of the arbitrariness of the resummation procedure, it would not
be correct to conclude that the reduction in the theoretical uncertainty
is proportional to the reduction in the scale dependence.

The final-state phase space includes integrations over the virtualities
$m_c^2x$ and $m_c^2y$ of the final-state gluons. Near zero virtuality,
one is in a region in which perturbation theory breaks down and
nonperturbative contributions may be important. We have identified this
region of low virtuality with contributions to the NRQCD matrix element
of a color-octet, ${}^3S_1$ four-fermion operator. The contribution of
this matrix element is of relative order $v^3$, and it is treated as an
uncertainty in our calculation.

A key ingredient in our analysis is the demonstration that there are no
contributions in which both gluons have small virtualities or in which
one gluon is exactly on shell and the other has a small virtuality. Such
contributions, if present, would not correspond to an NRQCD operator
matrix element and would be inconsistent with the NRQCD factorization
program. We demonstrate the absence of such contributions by writing the
integrals over the virtualities of the off-shell gluons as contour
integrals and deforming the contours out of the regions of small
virtuality.

One can understand this contour-deformation argument in terms of the
uncertainty principle. In the bubble-chain picture, away from the
kinematic limits of maximum gluon virtuality, the quarkonium decay
produces two energetic gluon jets, each with energy of order $m_c$.
Consider the case in which both gluons are virtual. That is, each gluon
jet contains two subjets that correspond to the constituents of a
vacuum-polarization bubble. The mass of each jet ranges up to a value of
order $m_c$. Now suppose that the jets are observed inclusively, that
is, without regard to their masses. Provided that neither jet has a mass
near the kinematic limit, the virtual-gluon masses are {\it
independently} uncertain by amounts of order $m_c$ and, according to the
uncertainty principle, the virtual gluons propagate over distances of
order $1/m_c$. Hence, the NRQCD four-fermion annihilation vertex is well
localized, in accordance with the NRQCD factorization formula. On the
other hand, if one restricts the gluon virtualities ({\it i.e.}, $x$ and
$y$) to be small, then the gluons propagate over long distances, of order
$1/(m_c\sqrt x)$ and $1/(m_c\sqrt y)$. Such propagation would delocalize
the annihilation vertex. Hence, contributions from small $x$ and $y$
must vanish in the inclusive rate. The absence of singularities in the
complex virtuality plane near the vacuum-polarization cut tells us that
there are no competing, nearly classical processes that would upset this
uncertainty-principle analysis of the bubble chains in isolation.

Note that this uncertainty-principle argument does not rule out
contributions in which a gluon is exactly on the mass shell (unless the
other gluon is virtual and at small virtuality). One should regard the
on-shell gluon not as propagating an infinite distance, but, rather, as
following a nearly classical trajectory. Since the momentum and energy
are uncertain by amounts of order $m_c$, the positions and times in the
trajectory are smeared by amounts of order $1/m_c$, and one again
reaches the conclusion that the NRQCD four-fermion annihilation vertex
is localized to within $1/m_c$ \cite{bbl}.

On the other hand, if the mass of one of the jets is near the kinematic
limit, then the mass (and uncertainty in mass) of the other jet is
constrained kinematically to be small, and the annihilation vertex
becomes delocalized. As we have seen, this kinematic region, in which
one virtuality is near the kinematic limit, yields long-distance
(delocalized) contributions that are absorbed into color-octet operator
matrix elements.

An interesting, and somewhat unexpected, feature of the bubble-chain
contributions is that their values depend on whether one carries out the
integration over the gluon virtualities before or after carrying out the
perturbation summation. The two orders of operations lead to results that
differ by an amount that has the characteristic structure
$\exp[-1/(\alpha_s\beta_0)]$ of a renormalon ambiguity. However, this
difference does not arise from a renormalon ambiguity since, in our
calculations, all of the renormalon contributions are absorbed into the
color-octet, ${}^3S_1$ matrix element. Rather, the difference arises from
the failure of the perturbation series to converge in the region of
small virtuality. The difference is associated with the presence of the
perturbative Landau pole in that region. We have argued, again using
contour integration, that one should integrate over the virtualities
before carrying out the perturbation summation. Then one can deform
contours out of the region of small virtuality and the resulting
perturbation series is reliable. On the other hand, there is no
justification for summing the perturbation series first: at sufficiently
small virtuality, one is outside the radius of convergence, and the
series is unreliable. We have argued that the order of operations in
which one carries out the integration over the gluon virtualities before
carrying out the perturbation summation is equivalent to the standard
Borel-transform method for evaluating the contribution of a bubble chain.

For $\alpha_s$ evaluated at the scale $\mu=2m_c$ and for typical values
of the cutoff $\delta$, the perturbation series, after integration over
the gluon virtualities, converges very slowly. Furthermore, the
integrands in high-order terms oscillate strongly and are difficult to
evaluate numerically. Therefore, we have devised a more efficient method
for numerical computation in which one carries out the perturbation
summation before the integrations over the gluon virtualities and then
makes corrections to account for the interchange of the order of limits.

Our numerical result for $R$ is much smaller than that which one obtains
by applying the BLM scale-setting method to the next-to-leading order
calculation.\footnote{If one regards the BLM procedure as an approximate
resummation of bubble contributions at one-loop accuracy in the bubble,
then it would be appropriate to evolve $\alpha_s$ from the nominal scale
$2m_c$ to the BLM scale by using the one-loop $\beta$ function. This
leads to $R^{\rm NLO}(\mu_{\rm BLM})=6.3\times 10^3$, which is smaller
than the value, obtained by 3-loop evolution, in Eq.~(\ref{R-BLM}).}
This may seem surprising, since the BLM method is designed to recover
the effects associated with the one-loop running of $\alpha_s$, to the
extent that the can be taken into account through a change of
renormalization scale. The effects of a change of scale in $\alpha_s$
can be expanded in a perturbation series. At one-loop accuracy in the
running of $\alpha_s$, one obtains the geometric series
\begin{equation}
\alpha_s(\mu')={\alpha_s(\mu)\over 1-2\alpha_s(\mu)\beta_0\ln(\mu/\mu')}
=\alpha_s(\mu)\sum_{n=0}^\infty [2\alpha_s(\mu)\beta_0
\ln (\mu/\mu')]^n.
\end{equation}
It is illuminating to compare this series with the perturbation
expansion for the contribution $G_1(\mu)$ [Eq.~(\ref{G1})] of a single
bubble chain, integrated over the gluon virtuality. In the NNA case we
have
\begin{eqnarray}
G_1(2m_c)=&&
1+1.91 \alpha_s(2m_c) +2.47 \alpha_s^2(2m_c) +0.97\alpha_s^3(2m_c)
-4.49\alpha_s^4(2m_c)\nonumber\\
&&-11.76\alpha_s^5(2m_c)+\cdots\,.
\label{G1p}
\end{eqnarray}
We see that the individual terms in the expansion (\ref{G1p}) are {\it
not} well approximated by those of a geometric series. Therefore, we
would not expect the BLM method to reproduce this result accurately. Of
course, before integration over the gluon virtuality, the bubble chain
{\it is} a geometric series [Eq.~(\ref{bub-prop})], with a ratio between
terms of $-i\Pi(x)$. However, in order for the bubble-chain
contributions, integrated over phase space, to be close to a geometric
series, we must have
\begin{equation}
\langle [\Pi(x)]^n \rangle \approx \langle \Pi(x)
\rangle^n\,,
\label{average}
\end{equation}
where the angular brackets denote averaging over phase space, with the
other factors in the integrand for $R$ as a weight. The relation
(\ref{average}) holds only if the integration over the virtuality $x$ is
highly peaked. Hence, we do not expect, in general, to obtain a
geometric perturbation series. We note that there are generalizations of
the BLM method that take into account higher-order deviations of the
perturbation series from a geometric series \cite{brodsky-lu} and that
may reproduce the effects computed in this paper.

Finally, we remark that the methods that we have developed in this paper
can be applied to a variety of processes to resum large final-state
corrections associated with the running of $\alpha_s$. The methods are
most directly applicable to computations in NRQCD, but with small
modifications, they are also applicable to computations in full QCD. The
principal change in the method that is required in the latter case is to
absorb nonperturbative contributions into QCD matrix elements, rather
than into NRQCD matrix elements. In general, there are contributions
that are associated with the running of $\alpha_s$ in initial-state
virtual gluons, as well. Such initial-state contributions also contain
nonperturbative pieces that must be absorbed into operator matrix
elements.\footnote{Initial-state contributions associated with the
running of $\alpha_s$ have been discussed generally in terms of the
Borel-transform method \cite{beneke} and also for the specific case of
NRQCD decays \cite{braaten-chen,bodwin-chen}.} However, the bubble
series begins at relative order $\alpha_s^2$, instead of relative order
$\alpha_s$.
%
%%%%%%%%%%%%%%%%%%%%%%%%%%%%%%%%%%%%%%%%%%%%%%%%%%%%%%%%%%%%%%%%%%%%%%%%%%

\acknowledgements

We wish to thank G.~Peter Lepage for a number of illuminating
discussions and, in particular, for suggesting the interpretation of the
color-octet contribution to the decay rate. We also thank Kent
Hornbostel for the use of his 3-loop program for evolving $\alpha_s$. We
are grateful to Eric Braaten for a number of suggestions to improve the
manuscript. Work in the High Energy Physics Division at Argonne National
Laboratory is supported by the U.~S.~Department of Energy, Division of
High Energy Physics, under Contract No.~W-31-109-ENG-38.

\vfill\eject

%%%%%%%%%%%%%%%%%%%% appendices %%%%%%%%%%%%%%%%%%%%%%%%%%%%%%%%%%%%%%%%%%
\appendix
\section{The Gluon Self-Energy in the Background-Field Gauge}
\label{app:background}%
%\appendix{gluon self energy}

At one-loop level, a gluon self-energy receives contributions from
gluon, ghost and quark bubbles.  The value of a gluon bubble depends on
the choice of gauge. In this Appendix, we describe the calculation of
the gluon self-energy in the background-field
gauge\cite{background-field}.

In the background-field-gauge method, the gauge field is separated into
an external field or background field, which appears only in external
lines of a Feynman diagram, and an internal field, which appears only in
loops of a Feynman diagram. The gauge of the internal field is fixed,
but that of the external field is not. Hence, the action is explicitly
gauge covariant with respect to the external field, and the
background-field-gauge method maintains explicit gauge covariance with
respect to the external field in each step of a calculation. In the
background-field gauge, all of the logarithmic dependence on the
renormalization scale that is associated with the running of the
coupling constant is contained in the gluon self-energy \cite{abbott}.

In the $R_\xi$ class of Lorentz-covariant background-field gauges, the
gauge of the internal fields is specified by the parameter $\xi$, with
$\xi=1$ corresponding to the Feynman gauge \cite{abbott}. In the
background-field $R_\xi$ gauges, the contribution of a gluon bubble to
the gluon self-energy is given by
\begin{equation}\label{bub:g}
\Pi_{\mu\nu}^{ab\,gluon}(k,\xi)=-{\frac{1}2}g_s^2\mu^{2\epsilon}
f^{acd}f^{bdc}\int{\frac{d^Dp}{(2\pi)^D}}
{\frac{N_{\mu\nu}(p,k,\xi)}{p^2(p+k)^2}} \,,
\end{equation}
where
\begin{eqnarray}
N_{\mu\nu}(p,k,\xi)&=&\left\{
(2p+k)_\mu g_{\alpha\beta}+2k_\alpha g_{\mu\beta}
-2k_\beta g_{\mu\alpha}
+{1-\xi \over \xi}\left[ (k + p)_\alpha g_{\mu\beta}
+p_\beta g_{\mu\alpha}\right] \right\}\nonumber \\
&\times& \left\{(2p+k)_\nu g_{\gamma\delta}+
2k_{\gamma}g_{\nu\delta}
-2k_{\delta}g_{\nu\gamma}
+{1-\xi \over \xi}\left[(k + p)_{\gamma}g_{\nu\delta}
+p_{\delta} g_{\nu\gamma} \right] \right\}\nonumber \\
&\times &\left[g^{\alpha\gamma}-(1-\xi)
\frac{(p+k)^{\alpha}(p +k)^{\gamma}}{(p+k)^2}
\right ]\left[g^{\beta\delta}-(1-\xi)
\frac{p^{\beta} p^{\delta}}{p^2}\right ]\,.
\label{N}
\end{eqnarray}
A straightforward calculation, using Feynman parameter integrals, gives
\begin{eqnarray}
\Pi_{\mu\nu}^{ab\,gluon}(k,\xi)&=&
-i\left(k^2 g_{\mu\nu}-k_\mu k_\nu \right)\delta^{ab}
{\alpha_s \over 2\pi}C_A
\left({4\pi \mu^2 \over - k^2-i\epsilon}\right)^\epsilon\nonumber \\
& \times & \left\{
\Gamma(\epsilon)\Gamma(1-\epsilon)
\left[{\Gamma(3-\epsilon) \over \Gamma(4-2\epsilon)}
-2{\Gamma(1-\epsilon)\over \Gamma(2-2\epsilon)}\right]\;
+{1 \over 2}(\xi^2 -1)\right\}\,.
\label{bub:g:int}
\end{eqnarray}

The contribution of a ghost bubble to the gluon self-energy in the
background-field gauge is
\begin{eqnarray}
\Pi_{\mu\nu}^{ab\,ghost}(k)&=&g_s^2\mu^{2\epsilon}
f^{acd}f^{bdc}\int{\frac{d^Dp}{(2\pi)^D}}\,
{\frac{(2p+k)_\mu (2p+k)_\nu}{p^2(p+k)^2}} \nonumber\\
&=& i\left(k^2 g_{\mu\nu}-k_\mu k_\nu \right)\delta^{ab}
{\alpha_s \over 2\pi }C_A
\left({4\pi\mu^2 \over - k^2-i\epsilon} \right)^\epsilon 
{\Gamma(\epsilon)\Gamma(1-\epsilon)
\Gamma(2-\epsilon) \over \Gamma(4-2\epsilon)}\,.
\label{bub:ghost}
\end{eqnarray}

The contribution of $n_f$ massless quarks to the gluon self-energy is
\begin{eqnarray}
\Pi_{\mu\nu}^{ab\,quark}(k)&=& -g_s^2 n_f \mu^{2\epsilon}
\mathrm{Tr}(T^a \,T^b)
\int {\frac{d^Dp}{(2\pi)^D}}\,
{\frac{\mathrm{Tr} \left[(\not\! p+\not\! k)\gamma_\mu
\not\! p\gamma_\nu\right]} {p^2\,(p+k)^2}}\nonumber\\
&=& -i n_f
\left(k^2 g_{\mu\nu}- k_\mu k_\nu \right)\delta^{ab}
{\alpha_s \over \pi}
\left({4\pi\mu^2 \over - k^2-i\epsilon} \right)^\epsilon 
{\Gamma(\epsilon)
\Gamma^2(2-\epsilon) \over \Gamma(4-2\epsilon)}\,.
\label{bub:quark}
\end{eqnarray}

Adding (\ref{bub:g:int}), (\ref{bub:ghost}) and (\ref{bub:quark}) and
subtracting the pole in $\epsilon$ that corresponds to an ultraviolet
divergence, along with an associated constant that depends on the
renormalization scheme, we obtain the expression for the renormalized
one-loop gluon self-energy:
\begin{equation}
\Pi_{\mu\nu}^{ab}(k,\xi)=i
\left(k^2 g_{\mu\nu}- k_\mu k_\nu \right)\delta^{ab}
\beta_0\,\alpha_s {1\over\epsilon}
\left(bx^{-\epsilon}-a\right)\,,
\label{vac-pol-bg}
\end{equation}
where
\begin{mathletters}
\begin{eqnarray}
b &=& {\Gamma(1+\epsilon)  \over \beta_0\pi}
\left[C_A {\Gamma^2(1-\epsilon) \over \Gamma(2-2 \epsilon)}
- \left({C_A\over 2}+n_f\right)
{\Gamma^2(2-\epsilon)\over \Gamma(4-2 \epsilon)}
-{C_A\epsilon\over 4\beta_0\pi}(\xi^2-1)\right]\,
\left({\pi\mu^2 \over -m_c^2-i\epsilon}\right)^\epsilon\nonumber\\
&=&\Biggl\{1+\left[-\gamma+\ln 4\pi+{1\over \beta_0\pi}
\left({67 \over 12}-{5\over 18}n_f-{3\over 4}(\xi^2-1)\right)
+\ln {\mu^2 \over 4 m_c^2}\right]\epsilon+O(\epsilon^2)
\Biggr\}e^{i\pi\epsilon}\,,
\end{eqnarray}
\begin{equation}
a=1+\left(-\gamma+\ln 4\pi+{5\over 3}+C\right)\epsilon+O(\epsilon^2)\,,
\end{equation}
\label{a-b-background}%
\end{mathletters}
$\beta_0=(33-2n_f)/(12\pi)$, $x\equiv p^2/(4 m_c^2)$,
and $C$ is a renormalization-scheme-dependent constant. In the
$\overline{\rm MS}$ scheme, $C=-5/3$. Now, in Eqs.~(\ref{vac-pol-bg}) and 
(\ref{a-b-background}),
$\epsilon$ plays the role of an infrared regulator ($\epsilon<0$).

%%%%%%%%%%%%%%%%%%%%%%%%%%%%%%%%

\section{Computation of a Single Bubble Chain}
\label{app:single-chain}%

In this Appendix, we compute the real and virtual contributions from a 
single bubble chain. We consider first the real contribution. From
Eq.~(\ref{one-gluon-total}), we have
\begin{equation}
G_{1R}={1\over \pi}
\sum_{n=1}^\infty \int_0^1 dx\,
\left(1-{1\over x}\right)
{\rm Im\,}\left[{1\over \epsilon}\alpha_s
\left(bx^{-\epsilon}-a\right)
\right]^{n}\,,  
\label{bub}
\end{equation}
where $\epsilon<0$ is an infrared regulator. The two terms in
parentheses have different infrared behaviors. The contribution of the
second term contains infrared divergences, which appear as poles in
$\epsilon $. The contribution of the first term is free of infrared
divergences, and, so, we can take $\epsilon = 0$ in the integrand. Thus,
the contribution of the first term can be written as
\begin{equation}\label{bub:Ra}
G_{1Ra}={\frac{1}\pi } \sum_{n=1}^\infty
\int_{0}^1  dx \,{\rm Im\,}
\Big[\alpha_s\beta_0
\left(d-\ln x + i\pi \right)
\Big]^n \,,
\end{equation}
where $d$ is defined in Eq.~(\ref{def-d}).

Now let's evaluate the second term in parentheses in Eq.~(\ref{bub}). It
can be written as
\begin{eqnarray}
G_{1Rb} &=&-{\frac{1}\pi }
\sum_{n=1}^\infty
{\frac{(\alpha_s\beta _0)^n} {\epsilon^{n}} }
\,{\rm Im\,} \int_0^1{\frac{d x}x}{(b x^{-\epsilon}-a)}^n\nonumber\\
&=&{\frac{1}\pi }\sum_{n=1}^\infty
{\frac{(\alpha_s\beta_0)^n}{\epsilon^{n+1}}}
\,{\rm Im\,}\int_0^1 dy\,b
{\frac{\left[\left({by-a}\right)^n-(-a)^n\right] }
{(by-a)-(-a)}}\nonumber\\
&=&{\frac{1}\pi}
\sum_{n=1}^\infty{\frac{(\alpha_s\beta_0)^n}
{\epsilon^{n+1}}}
\,{\rm Im \,}\int_0^1 dy \,b
\sum_{m=0}^{n-1}(-a)^{(n-m-1)}\left({by-a}\right)^m\,,
\label{bub:Rb}
\end{eqnarray}
where we have made a change of integration variable $y=x^{-\epsilon }$,
and, in the second line, we have used the fact that $a$ is real.
Carrying out the integration over $y$, we obtain
\begin{eqnarray}
G_{1Rb}&=&{\frac{1}\pi }\sum_{n=1}^\infty
{\frac{(\alpha_s\beta_0)^n}{\epsilon^{n+1}}}
\,{\rm Im\,}\sum_{m=1}^n{\frac{1}m} (-a)^{(n-m)}\left({b-a}\right)^m
\nonumber\\
&=&{\frac{1}\pi}
\sum_{n=1}^\infty
{\frac{(-\alpha_s\beta _0 a)^n}{\epsilon^{n+1}}}
\,{\rm Im\,}
\left[-\ln {b \over a }-{1 \over n+1 } \left({a-b \over a}
\right)^{n+1} + O(\epsilon^{n+2})\right]\nonumber \\
&=&-\sum_{n=1}^\infty
{\frac{(-\alpha_s\beta_0 a)^n}{\epsilon^n}}
+{1 \over \pi \alpha_s\beta_0 }\sum_{n=2}^\infty
{1 \over n }\,{\rm Im \,}
\left(\alpha_s\beta_0{b-a \over \epsilon}\right)^{n}+O(\epsilon)\,,
\label{bub:Rb-int}
\end{eqnarray}
where, in the last line, we have used the fact that the phase of $b$,
exact to all orders in $\epsilon$, is $\pi\epsilon$. In the last line of
Eq.~(\ref{bub:Rb-int}), the first term contains infrared divergences,
and, since $b-a\sim \epsilon$, the second term is infrared finite.

From Eq.~(\ref{one-chain-v}), we see that the virtual correction from a
single bubble chain is
\begin{equation}\label{bub:vir}
G_{V}=
\sum_{n=1}^\infty
{\frac{(-\alpha_s\beta_0 a)^n}
{\epsilon^{\,n}}}\,.
\end{equation}
The terms in Eq.~(\ref{bub:vir}) with $n\geq 1$ exactly cancel the the
infrared divergences in the first term of the last line of
Eq.~(\ref{bub:Rb-int}). The $n=0$ term of Eq.~(\ref{bub:vir}) and the
remainder of Eq.~(\ref{bub:Rb-int}) combine to give
\begin{equation}
G_{1Rb}+G_{V}={-1\over \pi \alpha_s\beta_0 }
{\rm Im \,} \ln \left[1 - \alpha_s\beta_0
{b-a \over \epsilon} \right]
={1 \over \pi \alpha_s\beta_0 }
\arctan {{ \pi \alpha_s\beta_0}\over
{1-\alpha_s\beta_0 d} }\,.
\label{bub:sum}
\end{equation}

Combining Eqs.~(\ref{bub:Rb}) and (\ref{bub:sum}), we obtain
\begin{eqnarray}
G_1&&=G_{1Ra}+G_{1Rb}+G_{V} \nonumber\\
&&={1 \over \pi \alpha_s\beta_0 }
\arctan {\pi \alpha_s\beta_0\over
1-\alpha_s\beta_0 d}
+{\frac{1}\pi} \sum_{n=1}^\infty
\int_{0}^1 dx\,{\rm Im\,}
\Big[\alpha_s\beta_0
(d-\ln x + i\pi)
\Big]^n  \,.
\label{width}
\end{eqnarray}

\section{Renormalon in the Short-Distance Coefficients}
\label{app:renormalon}%

In this Appendix, we compute the leading renormalon singularities in the
Borel transforms of the short-distance coefficients of $R$.\footnote{For
reviews of the properties of renormalons and Borel transforms, see
Refs.~\cite{beneke,bodwin-chen}.} We consider the
expressions for $G_1$ and $G_2$, Eqs.~(\ref{G1}) and (\ref{G2}), without
the cutoff (\ref{delta}). Once the cutoff (\ref{delta}) has been
imposed, the renormalon singularities are absent in $G_1$ and $G_2$.

The Borel transform of a function $f(\alpha_s)$ that has a perturbation
expansion
\begin{equation}
f(\alpha_s)=\sum_{n=0}^\infty a_n\alpha_s^n
\end{equation}
is defined by
\begin{equation}
B[f](u)=a_0\delta(u/\beta_0)+\sum_{n=1}^\infty {a_n\over (n-1)!}
(u/\beta_0)^{n-1}.
\end{equation}
The renormalon singularities in the Borel transform are those that arise
from the bubble-chain sum, and the leading singularity is the one that
appears at the smallest value of $u$.

Renormalon singularities can arise from the ultraviolet or infrared
regions of momentum integrals. Since, in $R$, the virtualities $x$ and
$y$ are bounded by unity, there are no ultraviolet renormalons. We have
argued (in Sec.~\ref{sec:single-chain}) that $G_1$ receives no
contributions from the region of small virtuality. Hence, it contains no
renormalons. We have also argued (in Sec.~\ref{sec:2-chains}) that $G_2$
receives no contributions in which both $x$ and $y$ are small.
Therefore, the renormalons in $G_2$, can arise only from the regions
$x\rightarrow 0$ and $y\sim 1$ or $y\rightarrow 0$ and $x\sim 1$.

We consider the region $x\rightarrow 0$ and $y\sim 1$ and multiply our
result by two to account for the contribution of the second region. In
this region, in the expression (\ref{G2}), we can approximate $g_n(y)$
by $g_n(1)$. Corrections to this yield behavior in $x$ that is
subleading as $x\rightarrow 0$. Furthermore, the $g_n(1)$ yield a
convergent series, and, so, we may carry out the Borel transform term by
term. From the relation
\begin{equation}
B[\alpha_s f](u)={1\over \beta_0}\int_0^u du'\, B[f](u')\,,
\label{borel-recurs}
\end{equation}
we see that the leading singularity in the Borel transform of $G_2$ must
come from the leading power of $\alpha_s$ in the terms $g_n(1)$: the
integration on the right side of Eq.~(\ref{borel-recurs}) weakens the
singularity as the power of $\alpha_s$ from the $g_n(1)$ increases.
Therefore, we retain only the leading term $g_1(1)=\alpha_s\beta_0$, to
obtain
\begin{eqnarray}
{1\over 2}B[G_2](u) &\sim & -\beta_0\int^1_0{dx\over x } 
(1-x) B[\alpha_s g](x,u)
      \int^1_{(1-\sqrt{x})^2} {dy \over y } (1-y) \nonumber \\
 &&    +\beta_0\int^1_0 {dx \over x } B[\alpha_s g](x,u) 
       \int_0^{(1-\sqrt{x})^2} {dy \over y }\,
  [f(x,y)-(1-x)(1-y)]\;,
\label{G2-ren}
\end{eqnarray}
where 
\begin{eqnarray}
B[\alpha_s g](x,u) &=&
  {1\over \pi}{\rm Im}\,\left(x^{-u}e^{du+i\pi u} \right)
  \nonumber \\
&=&x^{-u}e^{du}\,\sin(\pi u) \;. 
\label{gu}
\end{eqnarray}
An analysis of the integrations over $y$ yields the leading
contributions for $x$ small. Inserting these results into
Eq.~(\ref{G2-ren}), we have
\begin{equation}
{1\over 2}B[G_2](u) \sim  -\beta_0\int^1_0{dx\over x } 
(1-x) B[\alpha_s g](x,u)(2/3)x^{3/2}
+\beta_0\int^1_0 {dx \over x } B[\alpha_s g](x,u) 3x\ln x\,,
\label{y-integrated}
\end{equation}
where the first and second terms on the right side of
Eq.~(\ref{y-integrated}) correspond to the first and second terms,
respectively, on the right side of Eq.~(\ref{G2-ren}). Then, a
straightforward calculation gives the leading renormalon singularity for
each of the two terms in Eq.~(\ref{G2-ren}):
\begin{equation}\label{ren}
  B[G_2](u) \sim
  {\beta_0\over \pi} e^{du}\left[-{8\over 3}{\sin(\pi u)
  \over 3-2u} -6{\sin(\pi u) \over (1-u)^2} \right] \,,
\label{renorm-result}
\end{equation}
where the first and second terms on the right side of 
Eq.~(\ref{renorm-result}) correspond to the first and second terms,
respectively, on the right sides of Eqs.~(\ref{G2-ren}) and (\ref{G2}).

\vfill \eject

\end{document}